\documentclass{article}

\title{Conformal Transformation of the Schr\"{o}dinger Equation for the Harmonic Oscillator into a Simpler Form}
\author{Robert Ducharme}

\begin{document}
\maketitle

\centerline{151 Fairhills Dr., Ypsilanti, MI 48197}
\centerline{E-mail: ducharme01@comcast.net}

\begin{abstract}
The Schr\"{o}dinger equation and ladder operators for the harmonic oscillator are shown to simplify through the use of an isometric conformal transformation. These results are discussed in relation to the Bargmann representation. It is further demonstrated that harmonic interactions can be introduced into quantum mechanics as an imaginary component of time equivalent to adding the oscillator potential into the hamiltonian for the confined particle.
\end{abstract}

\section{Introduction}
Conformal mapping is often presented as a method for simplifying mathematical problems in two dimensions. Liouville's theorem \cite{DEB} in fact shows that higher dimensional conformal maps are possible but must be composed of translations, similarities, orthogonal transformations and inversions. The purpose of this paper is to present an application of a higher-dimensional conformal transformation in the simplification of the Schr\"{o}dinger equation for the three-dimensional harmonic oscillator. 

Suppose an oscillator of energy E consists of a single particle with a spatial displacement $x_i$ $(i=1,2,3)$ from the source of the potential confining it at time t. The goal of section 2 of this paper is to introduce an isometric conformal mapping of the form $z_i= x_i,s=t+\imath f(|x_i|,E)$ where $f$ is a real function and $\imath=\sqrt{-1}$. It is clear from inspection that this passive transformation does no more than introduce an imaginary shift in the time of the two related coordinate systems.

It is convenient to write the complex conjugate form of the $(z_i,s)$-coordinates as $(z_i^*,s^*)$ even though $z_i^*=z_i$ since $z_i^*$ and $z_i$ still belong to different coordinate systems. This distinction is shown to be most evident in the computation of the partial derivatives $\partial / \partial z_i$ and $\partial / \partial z_i^*$ from the chain rule of partial differentiation as these evaluate differently in their conjugate coordinate systems. One further topic to be introduced in section 2 is the Cauchy-Riemann equations that are necessary to determine if a function transformed into the $(z_i,s)$-coordinate system has well defined partial derivatives. 

The Schr\"{o}dinger equation for the three-dimensional harmonic oscillator is presented in section 3 alongside a complete set of eigensolutions. It is shown that both these results simplify in terms $(z_i,s)$-coordinates and that the eigensolutions are holomorphic. This argument is used to demonstrate that harmonic interactions may be introduced into the free-field Schr\"{o}dinger equation by adding an imaginary component to the world time of the particle. The imaginary part of time has a confining effect on the particle equivalent to including an oscillator potential in the hamiltonian.

In section 4, it is demonstrated that $\kappa \partial / \partial z_i$ and $-\kappa \partial / \partial z_i^*$ are respectively the operators for lowering and raising the eigenstates of the harmonic oscillator where $\kappa$ is a scaling constant. These results are compared to both the Bargmann \cite{VB} and conjugate form of the Bargmann representation \cite{RNA} .

\section{Conformal Mapping}
The task ahead is to present a isometric conformal transformation relating a real $(x_i,t)$-coordinate system and a complex $(z_i,s)$ coordinate system. This mapping is to be applied in section 3 to a harmonic oscillator consisting of a single particle of mass m and total energy E. It is convenient to express it in the form
\begin{equation} \label{eq: conftrans1}
z_{i} = x_{i}, \quad s = t - \imath \frac{ m\omega}{2 E}x^2 
\end{equation}
where $\omega$ is the spring constant of the oscillator . In the $(x_i,t)$-coordinate system, the particle has a spatial displacement $x_i$ from the source but shares the same world time t. Similarly, in the $(z_i,s)$-coordinate system, the particle has a dispacement $z_i = x_i$ from the source but shares the same complex time s. The isometric nature of the transformation therefore follows from the result $|z_i|=|x_i|$. It is also clear that the complex time $s$ is translated through an imaginary displacement $\imath \frac{ m\omega}{2 E}x^2$ from the real time $t$. 

In the application of complex coordinates to express physical problems, there is generally going to be both a complex and a complex conjugate coordinate representation for each individual problem. In the present case, the complex conjugate of eq. (\ref{eq: conftrans1}) is
\begin{equation} \label{eq: conftrans2}
z_{i}^* = x_{i}, \quad s^*= t + \imath \frac{m\omega}{2E}x^2 
\end{equation}
Naturally, there must also be inverse transformations mapping the complex and complex conjugate representations of the problem back into a single physical coordinate system. The inverses of the transformations (\ref{eq: conftrans1}) and (\ref{eq: conftrans2}) are readily shown to be
\begin{equation} \label{eq: inv_ict1}
x_{i} = z_{i}, \quad t = s + \imath \frac{m\omega}{2E}z^2 
\end{equation}
\begin{equation} \label{eq: inv_ict2}
x_{i} = z_{i}^*, \quad t = s^* - \imath \frac{m\omega}{2E}z^2 
\end{equation}
respectively.  

It is now interesting to investigate properties of derivatives with respect to complex 4-position coordinates. In particular, the chain rule of partial differentiation gives
\begin{equation}  \label{eq: complexDiff1}
\frac{\partial}{\partial s}  
= \frac{\partial t}{\partial s} \frac{\partial}{\partial t}
+ \frac{\partial x_{i}}{\partial s} \frac{\partial}{\partial x_{i}}
= \frac{\partial}{\partial t} 
\end{equation}
\begin{equation} \label{eq: complexDiff2}
\frac{\partial}{\partial s^*}  
= \frac{\partial t}{\partial s^*} \frac{\partial}{\partial t}
+ \frac{\partial x_{i}}{\partial s^*} \frac{\partial}{\partial x_{i}}
= \frac{\partial}{\partial t} 
\end{equation}
\begin{equation} \label{eq: complexDiff3}
\frac{\partial}{\partial z_i} 
= \frac{\partial x_i}{\partial z_i} \frac{\partial}{\partial x_i}
+ \frac{\partial t}{\partial z_i} \frac{\partial}{\partial t}
= \frac{\partial}{\partial x_i} + \imath \frac{m \omega x_i}{E} \frac{\partial}{\partial t}
\end{equation}
\begin{equation} \label{eq: complexDiff4}
\frac{\partial}{\partial z_i^*} 
= \frac{\partial x_i}{\partial z_i^*} \frac{\partial}{\partial x_i}
+ \frac{\partial t}{\partial z_i^*} \frac{\partial}{\partial t}
= \frac{\partial}{\partial x_i} - \imath \frac{m \omega x_i}{E} \frac{\partial}{\partial t}
\end{equation}
Note, eqs. (\ref{eq: complexDiff1}) and (\ref{eq: complexDiff3}) have been obtained using eq. (\ref{eq: inv_ict1});  eqs. (\ref{eq: complexDiff2}) and (\ref{eq: complexDiff4}) are based on eq. (\ref{eq: inv_ict2}). It has also been assumed in deriving eqs. (\ref{eq: complexDiff1}) through (\ref{eq: complexDiff4}) that
\begin{equation} \label{eq: complexDiff5}
\frac{\partial z_\nu}{\partial s}=\frac{\partial s}{\partial z_\mu} = \frac{\partial z_\nu^*}{\partial s^*}=\frac{\partial s^*}{\partial z_\mu^*}=0
\end{equation}
indicating that the coordinates $z_i$ and $s$ are independent of each other as are the complex conjugate coordinates $z_i^*$ and $s^*$. This assumption is readily validated using eqs. (\ref{eq: complexDiff1}) through (\ref{eq: complexDiff4}) to directly evaluate each of the derivatives in eq. (\ref{eq: complexDiff5}) in $(x_i,t)$-coordinates. 

In further consideration of eqs. (\ref{eq: conftrans1}), it is convenient to write $s=t+iy$ where $y = -(m \omega / 2E) x^2$. The requirement for a continuously differentiable function $\tau(s)=g(t,y)+ih(t,y)$ to be holomorphic is then for the real functions g and h to satisfy the set of Cauchy-Riemann equations
\begin{equation}
\frac{\partial g}{\partial y}=\frac{\partial h}{\partial t} \quad
\frac{\partial h}{\partial y}=-\frac{\partial g}{\partial t} 
\end{equation}
or equivalently
\begin{equation} \label{eq: creq}
\frac{\partial^2 \tau}{\partial t^2} + \frac{4E^2}{m^2 \omega^2} \frac{\partial^2 \tau}{\partial x^4} = 0
\end{equation}
It is thus concluded that a function $\psi(x_i,t)$ will also have an equivalent holomorphic form $\theta(z_i)\tau(s)$ in the complex $(z_\mu, s)$-coordinate system providing it is separable and $\tau$ satisfies eq. (\ref{eq: creq}). Here, it is understood that the domain of the Cauchy-Riemann equations in this problem is the complex plane containing s. The Cauchy-Riemann equations put no restriction at all on the form of the function $\theta(z_i)$ since $z_i$ and $s$ are independent coordinates and $z_i$ belongs to a real three-dimensional space.

\section{The Harmonic Oscillator}
The Schr\"{o}dinger equation determining the wavefunction $\psi(x_i, t)$ for a single particle confined in a 3-dimensional harmonic oscillator potential can be expressed in the form
\begin{equation} \label{eq: schrod1}
-\frac{\hbar^2}{2 m} \frac{\partial^2 \psi}{\partial x_i^2} + \frac{1}{2} m \omega^2 x^2\psi = E\psi
\end{equation}
where $\hbar$ is Planck's constant divided by 2$\pi$ and
\begin{equation} \label{eq: schrod2}
E\psi = \imath \hbar \frac{\partial \psi}{\partial t} 
\end{equation}
gives the total energy of the particle.

The solution \cite{DFL} to eqs. (\ref{eq: schrod1}) and (\ref{eq: schrod2}) takes the separable form
\begin{eqnarray} \label{eq: psi1} 
\psi(x_i,t) = \phi_{l_1}(x_1)\phi_{l_2}(x_2)\phi_{l_3}(x_3)\exp(-\imath Et / \hbar)
\end{eqnarray}
where
\begin{eqnarray} \label{eq: phi1} 
\phi_l(x_i) =   k_l H_{l}(\xi_i) \exp \left(-\frac{\xi^2}{2} \right)
\end{eqnarray}
$\xi_i=\sqrt{\frac{m \omega}{\hbar}}x_i$, $H_{l_j}$ are Hermite polynomials and $l_1,l_2,l_3$ are positive integers. The normalization constant is
\begin{equation}
k_l = \left( \frac{2 m \omega}{\hbar} \right)^{1/4}  \left( \frac{1}{2^l l!} \right)^{1/2}
\end{equation}

In developing the connection between complex $(z_i,s)$-coordinates and the quantum harmonic oscillator, eqs. (\ref{eq: complexDiff3}), (\ref{eq: complexDiff4}) and (\ref{eq: schrod2}) can be combined to give
\begin{equation} \label{eq: complexDiff6}
\frac{\partial}{\partial z_i} 
= \frac{\partial}{\partial x_i} + \frac{m \omega}{\hbar} x_i 
\end{equation}
\begin{equation} \label{eq: complexDiff7}
\frac{\partial}{\partial z_i^*} 
= \frac{\partial}{\partial x_i} - \frac{m \omega}{\hbar} x_i
\end{equation}
These results lead to the operator relationship
\begin{equation}\label{eq: qprop1}
 -\frac{\hbar^2}{2m}\frac{\partial}{\partial z_i^* \partial z_i} + \frac{3}{2}\hbar \omega  = -\frac{\hbar^2}{2m}\frac{\partial}{\partial x_i^2}+ \frac{1}{2} m \omega^2 x^2 
\end{equation}
enabling the Schr\"{o}dinger equation (\ref{eq: schrod1}) for the harmonic oscillator to be expressed in the concise form
\begin{equation}\label{eq: complexSchrod1}
-\frac{\hbar^2}{2m}\frac{\partial \psi}{\partial z_i^* \partial z_i} + \frac{3}{2}\hbar \omega \psi= E \psi
\end{equation}
It is also readily shown using eqs. (\ref{eq: complexDiff1}) and (\ref{eq: schrod2}) that
\begin{equation} \label{eq: complexSchrod2}
E\psi = \imath \frac{\partial \psi}{\partial s} 
\end{equation}
Eqs. (\ref{eq: complexSchrod1}) and (\ref{eq: complexSchrod2}) together, therefore, constitute a complete description of the quantum harmonic oscillator in terms of $(z_i,s)$-coordinates. On comparing eq. (\ref{eq: schrod1}) and (\ref{eq: complexSchrod1}), it is clear that the harmonic oscillator potential term in the original Schr\"{o}dinger equation is replaced by a constant term in complex coordinates.

The oscillator function (\ref{eq: psi1}) is readily transformed into  $(z_i,s)$-coordinates using eqs. (\ref{eq: conftrans1}) to give
\begin{eqnarray} \label{eq: psi3} 
\psi(z_i,s) = \theta_{l_1}(z_1)\theta_{l_2}(z_2)\theta_{l_3}(z_3)\tau(s)
\end{eqnarray}
where
\begin{eqnarray} \label{eq: psi4} 
\theta_l(z_i) =  k_l H_{l}(\zeta_i), \quad \tau(s) = \exp(-\imath Es / \hbar)
\end{eqnarray}
and $\zeta_i = \sqrt{\frac{m \omega}{\hbar}}z_i$. It is notable that eq. (\ref{eq: psi1}) and (\ref{eq: psi3}) are similar except that eq. (\ref{eq: psi3}) does not contain a gaussian term. It is also notable that $\tau(s)$ is a continuously differentiable solution of eq.(\ref{eq: creq}) thus demonstrating that the oscillator function $\psi(z_i, s)$ is holomorphic.

In consideration of the foregoing arguments, it is of interest that eqs. (\ref{eq: conftrans1}) reduces to the form $z_i=x_i, s=t$ on setting $\omega=0$. It also apparant that eq. (\ref{eq: complexSchrod1}) reduces to the free field form of the Schr\"{o}dinger equation under these same conditions. The converse of this argument is that harmonic interactions may be introduced into the free-field Schr\"{o}dinger equation through the replacement $t \rightarrow t - \imath \frac{ m\omega}{2 E}x^2$ exactly equivalent to the more usual approach of adding the oscillator potential into the hamiltonian for the oscillator.

\section{Ladder Operators}
As is well known, the Schr\"{o}dinger equation (\ref{eq: schrod1}) can be simplified in terms of the non-relativistic raising and lowering operators: 
\begin{equation} \label{lower_nr}
\hat{a}_i = \sqrt{\frac{\hbar}{2 m\omega}} \frac{\partial}{\partial x_i} + \sqrt{\frac{m\omega}{2 \hbar}} x_i 
\end{equation}
\begin{equation} \label{raise_nr}
\hat{a}_i^\dag = -\sqrt{\frac{\hbar}{2 m\omega}} \frac{\partial}{\partial x_i} + \sqrt{\frac{m\omega}{2 \hbar}} x_i
\end{equation}
to give
\begin{equation} \label{schrod3}
\hat{a}_i^\dag \hat{a}_i \psi + \frac{3}{2}\psi = \frac{E}{\omega \hbar} \psi
\end{equation}
On comparing eq. (\ref{eq: complexDiff6}) and (\ref{eq: complexDiff7}) with eq. (\ref{lower_nr}) and (\ref{raise_nr}), it can be seen that
\begin{equation}  \label{ladder_nr}
\hat{a}_i = \frac{1}{\sqrt{2}}\frac{\partial}{\partial \zeta_i}, \quad \hat{a}_i^\dag = -\frac{1}{\sqrt{2}}\frac{\partial}{\partial \zeta_i^*}
\end{equation} 
It is clear therefore that raising and lowering operators have a more concise representation in complex $(z_i,s)$-coordinates than in real $(x_i,t)$-coordinates.

Applying ladder operators to the eigenfunctions for the harmonic oscillator gives
\begin{equation} \label{eq: lower1}
\hat{a}_1\psi(l_1, l_2, l_3) = \sqrt{l_1} \psi(l_1-1, l_2, l_3)
\end{equation}
\begin{equation} \label{eq: lower2}
\hat{a}_2\psi(l_1, l_2, l_3) = \sqrt{l_2} \psi(l_1, l_2-1, l_3)
\end{equation}
\begin{equation} \label{eq: lower3}
\hat{a}_3\psi(l_1, l_2, l_3) = \sqrt{l_3} \psi(l_1, l_2, l_3-1)
\end{equation}
for lowering the state of the oscillator, alongside the conditions
\begin{equation} \label{eq: raise1}
\hat{a}_1^\dag \psi(l_1, l_2, l_3)  = \sqrt{l_1+1} \psi(l_1+1, l_2, l_3)
\end{equation}
\begin{equation} \label{eq: raise2}
\hat{a}_2^\dag \psi(l_1, l_2, l_3)  = \sqrt{l_2+1} \psi(l_1, l_2+1, l_3)
\end{equation}
\begin{equation} \label{eq: raise3}
\hat{a}_3^\dag \psi(l_1, l_2, l_3)  = \sqrt{l_3+1} \psi(l_1, l_2, l_3+1)
\end{equation}
for raising it. 

It is interesting now to compare the results of conformal transformation in this paper with two other complex formulations of the quantum harmonic oscillator. These other formulations are based on integral transformations. One of these is the Bargmann representation obtained using the Segal-Bargmann transformation:
\begin{equation} \label{sb_trans}
\theta_{l}(a_i) = \pi^{-1/4}\int \phi_{l}(\xi_i) \exp \left(\frac{-\xi_i^2-a_i^2}{2} \right) \exp \left( -\sqrt{2} a_i \xi_i \right) d\xi_i
\end{equation}
The other is the conjugate of the Bargmann representation given by
\begin{equation} \label{conjugate_trans}
\theta_{l}(b_i) = \pi^{-1/4}\int \theta_{l}(a_i)\exp \left( - a_i b_i \right) da_i
\end{equation}
The results of the transformations (\ref{sb_trans}) and (\ref{conjugate_trans}) are presented in table 1 including both ladder operators and eigenfunctions. The ladder operators can be inserted into eq. (\ref{schrod3}) and the eigenfunctions validated as the solution.

\begin{table}[h]
\centering

\begin{tabular}{|c|c|c|c|}
\hline
Space & $\hat{a}_i$ & $\hat{a}_i^\dag $ & $\theta_{l}$ \\ \hline
Bargmann & $\partial / \partial a_i$ &$a_i$ & $a_i^l / \sqrt{l!}$ \\ \hline
Conjugate & $b_i$ & -$\partial / \partial b_i$ & $\sqrt{l!} / b_i^{l+1}$ \\ \hline
Conformal & $2^{-1 / 2} \partial / \partial \zeta_i$ & -$2^{-1 / 2} \partial / \partial \zeta_i^*$ & $k_l H_l(\zeta_i)$\\ \hline
\end{tabular}
\caption{Ladder operators and eigenfunctions for the harmonic oscillator in different complex spaces.}
\label{tab:}
\end{table}

\section{Concluding Remarks}
It has been shown the mathematical description of the non-relativistic quantum harmonic oscillator can be simplified through the use of a conformal transformation. In the transformed coordinate system, time is a complex quantity. The real part of this complex time is the world time; the imaginary part represents the harmonic interaction confining the particle.

\end{document}